\documentclass[11pt,a4paper,fleqn]{article}

\usepackage{amsmath,amssymb,mathrsfs,amsthm,enumerate,cite}

\newcommand{\vspacebefore}{\raisebox{0ex}[2.5ex][0ex]{\null}}
\newcommand{\p}{\partial}
\newcounter{tbn}

\newcommand{\Equiv}{\sim}

\newcommand{\CV}{\mathop{\rm CV}\nolimits}
\newcommand{\CL}{\mathop{\rm CL}\nolimits}
\newcommand{\Ch}{\mathop{\rm Ch}\nolimits}
\newcommand{\Eop}{\mathop{\sf E}}
\newcommand{\Fder}{\mathop{\sf D}}
\newcommand{\Lop}{\mathop{\sf L}}

\newtheorem{theorem}{Theorem}

\newtheorem{corollary}{Corollary}
\newtheorem{lemma}{Lemma}
{\theoremstyle{definition}
\newtheorem{definition}{Definition}
\newtheorem{note}{Note}
\newtheorem*{note*}{Note}
\newtheorem{proposition}{Proposition}
}

\begin{document}
\begin{center}
{\LARGE\bf Conservation Laws of Multidimensional Diffusion--Convection Equations\\[1.5ex]}
{\large Nataliya~M.~Ivanova\\[1.7ex]}
\footnotesize
Institute of Mathematics of National Academy of Sciences of Ukraine, \\
3, Tereshchenkivska Str., Kyiv-4, 01601, Ukraine\\
Department of Mathematics, UBC, Vancouver, BC, V6T 1Z2, Canada\\
\vspace{1em}
E-mail: ivanova@imath.kiev.ua
\end{center}

{\vspace{7mm}\par\noindent\hspace*{8mm}\parbox{145mm}{\small
All possible linearly independent local conservation laws for $n$-dimensional diffusion--convection equations
$u_t=(A(u))_{ii}+(B^i(u))_i$ were constructed using the direct method and the composite variational principle.
Application of the method of classification of conservation laws with respect to the group of point transformations
[R.O.~Popovych, N.M.~Ivanova, {\it J. Math. Phys.}, 2005, V.46, 043502 (math-ph/0407008)]
allows us to formulate the result in explicit closed form.
Action of the symmetry groups on the conservation laws of diffusion equations is investigated and generating sets of conservation laws
are constructed.
}\par\vspace{7mm}}

\section{Introduction}

One of the most important applications of group analysis is the construction of conservation laws of (systems of) differential equations.
They provide information on the basic properties of solutions of differential equations.
The famous laws of conservation of energy, linear momentum and angular momentum are important
tools for solving many problems arising in mathematical physics.
 Knowledge of conservation laws is important for numerical integration
of PDEs. Investigation of conservation laws of Korteweg--de Vries equation became a starting point of discovery
of new approaches to integration of PDEs (such as Miura transformations, Lax pairs, inverse scattering, bi-Hamiltonian structures etc.)
Existence of the `sufficient number' of conservation laws of (systems of) PDEs is a reliable indicator of their possible integrability.

In view of the generalized Noether theorem, there exists a one-to-one correspondence between
the non-trivial generalized variational symmetries of some variational functional and the non-trivial conservation laws
of the associated Euler--Lagrange equations,
and any such symmetry is a generalized symmetry of the Euler--Lagrange equations.
Thus, e.g., conservation of energy arises from the invariance of the corresponding variational problem
with respect to the group of time translation.
The Noether approach reduces construction of conservation laws to finding symmetries
for which there exist a number of well-developed methods.
However, this approach can be applied only to Euler--Lagrange equations that form normal systems and admit
symmetry groups satisfying an additional ``variational'' property of leaving the variational integral invariant in
some sense~\cite{Olver1986}. The latter requirements lead to restriction of class of systems that could be
investigated in such way.

At~the same time, the definition of conservation laws itself yields a method of finding conservation laws.
This method is called {\em direct}, and it is possible to distinguish four its versions,
depending on the way of taking into account systems under investigation.
(See e.g.~\cite{Anco&Bluman2002a,Anco&Bluman2002b,Popovych&Ivanova2005CLs,Wolf2002,Olver1986}
for description and detailed comparison of different methods of finding conservation laws.)
In the present paper we use its most direct version based on the immediate solving of determining equations for
conserved vectors on the solution manifolds of investigated systems.

Let us note that there exist other approaches for construction of conservation laws
which differ from the Noether or direct ones,
are based on exploitation of symmetry properties of differential equations
and can be applied  to non-Lagrangian systems.
Thus, W.I.~Fushchych and A.G.~Nikitin~\cite{Fushchych&Nikitin1994} proposed to directly calculate bilinear combinations
of solutions of motion equations, which are conserved in time by virtue of symmetries of these equations.

Another useful approach for construction of conservation laws of non-Lagrangian systems
is so-called {\em composite variational principle}~\cite{Vanberg1956,Atherton&Homsy1975}.
Composite principle means that in addition to the original variables of a given system, one should introduce a set of adjoint variables
in order to obtain a system of Euler--Lagrange equations for some variational functional.
To the best of our knowledge such method was firstly developed and applyed
for finding conservation laws of the linear one-dimensional heat equation~\cite{Atherton&Homsy1975}.
Recently this approach was rediscovered and used by N.H.~Ibragimov and T.~Kolsrud for investigation of conservation laws
of one-dimensional diffusion equations~\cite{Ibragimov&Kolsrud2004}.

The aim of this work is to construct all possible linearly independent local conservation laws
of $n$-dimensional diffusion--convection equations of the form
\begin{equation} \label{eq_DifConvMultiDim}
u_t=(A(u))_{ii}+(B^i(u))_i,
\end{equation}
where $A(u)$ and $B^i(u)$ are arbitrary smooth functions, $A_u\ne0$,
$u=u(x)=u(x^1,\ldots,x^n)$, $u_i={\p u}/{\p x^i}$, $u_{ij}={\p^2 u}/{\p x^i\p x^j}$.
Here and below we use the summation convention for repeated indices, $i=1,\ldots, n$.
This equation, often called the Richard's equation, arises naturally in certain physical applications.
Thus, for example, superdiffusivities of this type have been proposed~\cite{Gennes1983} as a model for long-range Van der Waals interactions in thin
films spreading on solid surfaces. This equation also appears in the study of
cellular automata and interacting particle systems with self-organized criticality (see \cite{Chayes&Osher&Ralston1993} and references therein).
It describes a model of water flow in unsaturated soil~\cite{Richard's1931}.

The main tool of our investigation is the notion of equivalence of conservation laws with respect to equivalence groups
which was introduced in~\cite{Popovych&Ivanova2005CLs}.
Investigation of such equivalences allows us to simplify essentially technical calculations and to obtain results
in closed explicit form. 
Note that for wide classes of nonlinear equations, this is the only way
to obtain the complete description of spaces of conservation laws
(see, e.g.,~\cite{Ivanova&Popovych&Sophocleous2004,Popovych&Ivanova2005CLs} for detailed analysis of application of such equivalences
for finding conservation laws of different classes of one-dimensional diffusion--convection equations).

Conservation laws of equations~\eqref{eq_DifConvMultiDim} were studied firstly for its one-dimensional case.
Thus, A.H.~Kara and F.M.~Mahomed~\cite{Kara&Mahomed2002} constructed
the~first-order local conservation laws of equations~\eqref{eq_DifConvMultiDim} with $n=1$
and found the bases of such conservation laws with respect to the corresponding symmetry groups.
In~\cite{Ivanova2004,Popovych&Ivanova2005CLs}
the complete classification of local and potential conservation laws for one-dimensional equations~\eqref{eq_DifConvMultiDim} with respect
to the equivalence group is presented.
V.A.~Dorodnitsyn and S.R.~Svirshchevskii~\cite{Dorodnitsyn&Svirshchevskii1983}
(see also~\cite[Chapter~10]{Ibragimov1994V1})
completely investigated the local conservation laws for one-dimensional reaction--diffusion equations $u_t=(A(u))_{xx}+C(u)$
having non-empty intersection with class~\eqref{eq_DifConvMultiDim}.
Conservation laws of multidimensional reaction--diffusion equations $u_t=\Delta(A(u))+f(u)$
were found by Y.R.~Romanovsky~\cite{Romanovsky1989}, who completely studied conservation laws for
case $A(u)=u^m$, and R.M.~Cherniha~\cite{Cherniha&King2005}, who considered the cases of power and exponential nonlinearities.

In the present work we briefly review the necessary theoretical background including the notions of equivalence of conservation laws with respect
to groups of equivalence and symmetry transformations (Section 2) introduced in our recent work~\cite{Popovych&Ivanova2005CLs},
construct the complete equivalence group and exhaustively classify,
with respect to this group, the local conservation laws of equations~\eqref{eq_DifConvMultiDim} (Section 3).
Then, in Section 4, we apply the composite variational principle to the subclass of diffusion equations with zeroth convectivities,
investigate the classical Lie symmetries of the obtained system of Euler--Lagrange equations, and use them
to construct conservation laws of the systems by means of Noether approach. In such a way we found a set of local
and nonlocal conservation laws for the diffusion equations, where the non-local variable is defined by the associated equation
for the adjoint symmetry. Action of symmetry groups on the conservation laws of diffusion equations is investigated
and generating sets of conservation laws are found in Section~4.

\section{Theoretical Background}\label{sec:def}

Let~$\mathcal{L}$ be a system~$L(x,u_{(\rho)})=0$ of $l$ differential equations $L^1=0$, \ldots, $L^l=0$
for $m$ unknown functions $u=(u^1,\ldots,u^m)$
of $n$ independent variables $x=(x^1,\ldots,x^n).$
Here $u_{(\rho)}$ denotes the set of all the derivatives of the functions $u$ with respect to $x$
of order not greater than~$\rho$, including $u$ as the derivatives of the zero order.
Let $\mathcal{L}_{(k)}$ denote the set of all algebraically independent differential consequences
that have, as differential equations, orders not greater than $k$. We identify~$\mathcal{L}_{(k)}$ with
the manifold determined by~$\mathcal{L}_{(k)}$ in the jet space~$J^{(k)}$.

\begin{definition}\label{def.conserved.vector}
A {\em conserved vector} of the system~$\mathcal{L}$ is
an $n$-tuple $F=(F^1(x,u_{(r)}),\ldots,F^n(x,u_{(r)}))$ for which the divergence ${\rm Div}\,F:=D_iF^i$
vanishes for all solutions of~$\mathcal{L}$ (i.e., ${\rm Div}F\bigl|_\mathcal{L}=0$).
\end{definition}

In definition~\ref{def.conserved.vector} and below
$D_i=D_{x^i}$ denotes the operator of total differentiation with respect to the variable~$x^i$.
The notation~$V\bigl|_\mathcal{L}$ means that values of $V$ are considered
only on solutions of the system~$\mathcal{L}$.

\begin{definition}
A conserved vector $F$ is called {\em trivial} if $F^i=\hat F^i+\check F^i,$ $i=\overline{1,n},$
where $\hat F^i$ and $\check F^i$ are, likewise $F^i$, functions of $x$ and derivatives of $u$
(i.e., differential functions),
$\hat F^i$ vanish on the solutions of~$\mathcal L$ and the $n$-tuple $\check F=(\check F^1,\ldots,\check F^n)$
is a null divergence (i.e., its divergence vanishes identically).
\end{definition}

\begin{definition}\label{DefinitionOfConsVectorEquivalence}
Two conserved vectors $F$ and $F'$ are called {\em equivalent} if
the vector-function $F'-F$ is a trivial conserved vector.
\end{definition}

For any system~$\mathcal{L}$ of differential equations the set~$\CV(\mathcal{L})$ of conserved vectors is a linear space,
and the subset~$\CV_0(\mathcal{L})$ of trivial conserved vectors is a linear subspace in~$\CV(\mathcal{L})$.
The factor space~$\CL(\mathcal{L})=\CV(\mathcal{L})/\CV_0(\mathcal{L})$
coincides with the set of equivalence classes of~$\CV(\mathcal{L})$ with respect to the equivalence relation adduced in
definition~\ref{DefinitionOfConsVectorEquivalence}.

\begin{definition}\label{DefinitionOfConsLaws}
The elements of~$\CL(\mathcal{L})$ are called {\em conservation laws} of the system~$\mathcal{L}$,
and the whole factor space~$\CL(\mathcal{L})$ is called {\em the space of conservation laws} of~$\mathcal{L}$.
\end{definition}

That is why we assume description of the set of conservation laws
as finding~$\CL(\mathcal{L})$ that is equivalent to construction of either a basis if
$\dim \CL(\mathcal{L})<\infty$ or a system of generatrices in the infinite dimensional case.
We will additionally identify elements from~$\CL(\mathcal{L})$ with their representatives
in~$\CV(\mathcal{L})$. Namely, we will understand a conservation law as a divergence expression $D_iF^i=0$
vanishing identically for all solutions of~$\mathcal{L}$.
In contrast to the order $r_F$ of a conserved vector~$F$ as the maximal order of derivatives explicitly appearing in~$F$,
the {\em order of a conservation law} as an element~$\cal F$ from~$\CL(\mathcal{L})$
is called $\min\{r_F\,|\,F\in\CV(\mathcal{L})\ \mbox{corresponds to}\ {\cal F}\}$.
Under the linear dependence of conservation laws, we understand the linear dependence of them as elements of~$\CL(\mathcal{L})$.

\begin{definition}\label{def.conservation.law.dependence}
Conservation laws of a system~$\mathcal{L}$ are called {\em linearly dependent} if
there exists their linear combination having a trivial conserved vector.
\end{definition}


Let the system~$\cal L$ be totally nondegenerate~\cite{Olver1986}.
Then application of the Hadamard lemma to the definition of conservation law and integrating by parts imply that
up to the equivalence relation of conserved vectors the left-hand side of any conservation law of~$\mathcal L$ can be always represented
as a linear combination of the left-hand sides of independent equations from $\mathcal L$
with coefficients~$\lambda^\mu$ being functions on a suitable jet space~$J^{(k)}$:
\begin{equation}\label{CharFormOfConsLaw}
\mathop{\rm Div}\nolimits F=\lambda^\mu L^\mu.
\end{equation}
Here the order~$k$ is determined by~$\mathcal L$ and the allowable order of conservation laws,
$\mu=\overline{1,l}$.

\begin{definition}\label{DefCharForm}
Formula~\eqref{CharFormOfConsLaw} and the $l$-tuple $\lambda=(\lambda^1,\ldots,\lambda^l)$
are called the {\it characteristic form} and the {\it characteristic}
of the conservation law~$\mathop{\rm Div}\nolimits F=0$ correspondingly.
\end{definition}

The characteristic~$\lambda$ is {\em trivial} if it vanishes for all solutions of $\cal L$.
Since $\cal L$ is nondegenerate, the characteristics~$\lambda$ and~$\tilde\lambda$ satisfy~\eqref{CharFormOfConsLaw}
for the same~$F$ and, therefore, are called {\em equivalent}
iff $\lambda-\tilde\lambda$ is a trivial characteristic.
Similarly to conserved vectors, the set~$\Ch(\mathcal{L})$ of characteristics
corresponding to conservation laws of the system~$\cal L$ is a linear space,
and the subset~$\Ch_0(\mathcal{L})$ of trivial characteristics is a linear subspace in~$\Ch(\mathcal{L})$.
The factor space~$\Ch_{\rm f}(\mathcal{L})=\Ch(\mathcal{L})/\Ch_0(\mathcal{L})$
coincides with the set of equivalence classes of~$\Ch(\mathcal{L})$
with respect to the above characteristic equivalence relation.

Using properties of total divergences, we can eliminate the conserved vector~$F$ from~\eqref{CharFormOfConsLaw}
and obtain a condition for the characteristic~$\lambda$ only.
Namely, a differential function~$f$ is a total divergence, i.e. $f=\mathop{\rm Div} F$
for some $n$-tuple~$F$ of differential functions iff $\Eop(f)=0$.
Hereafter the Euler operator~$\Eop=(\Eop^1,\ldots, \Eop^m)$ is the $m$-tuple of differential operators
\[
{\Eop}^a=(-D)^\alpha\p_{u^a_\alpha}, \quad a=\overline{1,m},
\]
where
$\alpha=(\alpha_1,\ldots,\alpha_n)$ runs the multi-indices set ($\alpha_i\!\in\!\mathbb{N}\cup\{0\}$),
$(-D)^\alpha=(-D_1)^{\alpha_1}\ldots(-D_m)^{\alpha_m}$.
Therefore, action of the Euler operator on~\eqref{CharFormOfConsLaw}
results to the equation
\begin{equation}\label{NSCondOnChar}
\Eop(\lambda^\mu L^\mu)={\Fder}_\lambda^*(L)+{\Fder}_L^*(\lambda)=0,
\end{equation}
which is a necessary and sufficient condition on characteristics of conservation laws for the system~$\mathcal{L}$.
The matrix differential operators~${\Fder}_\lambda^*$ and~${\Fder}_L^*$ are the adjoints of
the Fr\'echet derivatives~${\Fder}_\lambda^{\phantom{*}}$ and~${\Fder}_L^{\phantom{*}}$, i.e.,
\[
{\Fder}_\lambda^*(L)=\left((-D)^\alpha\left( \dfrac{\p\lambda^\mu}{\p u^a_\alpha}L^\mu\right)\right), \qquad
{\Fder}_L^*(\lambda)=\left((-D)^\alpha\left( \dfrac{\p L^\mu}{\p u^a_\alpha}\lambda^\mu\right)\right).
\]
Since ${\Fder}_\lambda^*(L)=0$ automatically on solutions of~$\mathcal{L}$ then
equation~\eqref{NSCondOnChar} implies a necessary condition for $\lambda$ to belong to~$\Ch(\mathcal{L})$:
\begin{equation}\label{NCondOnChar}
{\Fder}_L^*(\lambda)\bigl|_{\mathcal{L}}=0.
\end{equation}
Condition~\eqref{NCondOnChar} can be considered as adjoint to the criteria
${\Fder}_L^{\phantom{*}}(\eta)\bigl|_{\mathcal{L}}=0$ for infinitesimal invariance of $\mathcal{L}$
with respect to evolutionary vector field having the characteristic~$\eta=(\eta^1,\ldots,\eta^m)$.
That is why solutions of~\eqref{NCondOnChar} are sometimes called as
{\em cosymmetries}~\cite{Blaszak1998} or {\em adjoint symmetries}~\cite{Anco&Bluman2002b}.

Let now~$\mathcal{L}$ be a system of Euler--Lagrange equations $\Eop {\Lop}=0$ for the variational functional
$\mathscr{L}=\int \Lop(x,u_{(\rho)})dx$.
\begin{definition}\label{DefinitionOfVarInvar}
Generalized vector field $Q=\xi^i\p_i+\eta^a\p_{u^a}$ is called {\em variational symmetry} of functional $\mathscr{L}=\int \Lop(x,u_{(\rho)})dx$
iff there exists a tuple of $m$ differential functions $(B^1,\ldots,B^m)$, such that
$Q_{(\rho)}(\Lop)+\Lop D_i\xi^i=D_iB^i$ for all $x$ and $u$.
\end{definition}

It can be proved~\cite{Olver1986} that any variational symmetry of the variational problem is a symmetry
of the corresponding Euler--Lagrange equations. (In general, the inverse statement is not true.)
Thus, to find all variational symmetries of the given variational functional, it is enough to construct symmetries
of the corresponding Euler--Lagrange equations and then to check additional criterion of variational invariance
given in definition~\ref{DefinitionOfVarInvar}.

Variational symmetry $Q$ is {\em trivial} if its characteristic
$Q[u]=(\eta^1-\xi^iu^1_i,\ldots,\eta^m-\xi^iu^m_i)$
vanishes on the solution manifold of the Euler--Lagrange equations.
Two variational symmetries $Q^1$ and $Q^2$ are called {\em equivalent} if $Q^1-Q^2$ is a trivial symmetry.

There exists one-to-one correspondence between the non-trivial generalized variational symmetries of the variational functional
and the non-trivial conservation laws of the associated Euler--Lagrange equations,
and this correspondence is established by the generalized Noether theorem~\cite{Olver1986}:
\begin{theorem}
Generalized vector field $Q$ determines the group of variational symmetries of functional $\mathscr{L}=\int \Lop dx$ iff
its characteristic is a characteristic of a conservation law of the associated Euler--Lagrange equations $\Eop(\Lop)=0$.
In particular, if $\mathscr{L}$ is a nondegenerate variational problem, there exists a one-to-one correspondence between
the equivalence classes of nontrivial conservation laws of the Euler--Lagrange equations and equivalence classes
of variational symmetries of the functional.
\end{theorem}


In most of cases to describe completely the space of conservation laws of (a class of) systems of differential equations,
one should take into account symmetry transformations of a system
or equivalence transformations of a whole class of systems~\cite{Popovych&Ivanova2005CLs}.

\begin{proposition}
Any point transformation~$g$ maps a class of equations in the conserved form into itself.
More exactly, the transformation~$g$: $\tilde x=x_g(x,u)$, $\tilde u=u_g(x,u)$ prolonged to the jet space~$J^{(r)}$
transforms the equation $D_iF^i=0$ to the equation $D_iF^i_g=0$. The transformed conserved vector~$F_g$ is determined
by the formula
\begin{equation}\label{eq.tr.var.cons.law}
F_g^i(\tilde x,\tilde u_{(r)})=\frac{D_{x_j}\tilde x_i}{|D_x\tilde x|}\,F^j(x,u_{(r)}),
\quad\mbox{i.e.}\quad
F_g(\tilde x,\tilde u_{(r)})=\frac{1}{|D_x\tilde x|}(D_x\tilde x)F(x,u_{(r)})
\end{equation}
in the matrix notions. Here $|D_x\tilde x|$ is the determinant of the matrix $D_x\tilde x=(D_{x_j}\tilde x_i)$.
\end{proposition}

\begin{note}
In the case of one dependent variable ($m=1$) $g$ can be a contact transformation:
$\tilde x=x_g(x,u_{(1)})$, $\tilde u_{(1)}=u_{g(1)}(x,u_{(1)})$.
Similar notes are also true for the statements given below.
\end{note}

\begin{definition}
Let $G$ be a symmetry group of the system~$\mathcal{L}$.
Two conservation laws with the conserved vectors $F$ and $F'$ are called {\em $G$-equivalent} if
there exists a transformation $g\in G$ such that the conserved vectors $F_g$ and $F'$
are equivalent in the sense of definition~\ref{DefinitionOfConsVectorEquivalence}.
\end{definition}

\begin{proposition}
If system~$\mathcal{L}$ admits a one-parameter group of transformations then the infinitesimal generator
$Q=\xi^i\p_i+\eta^a\p_{u^a}$
of this group can be used for construction of new conservation laws from known ones.
Namely, differentiating equation~(\ref{eq.tr.var.cons.law})
with respect to the parameter $\varepsilon$ and taking the value $\varepsilon=0$,
we obtain the new conserved vector
\begin{equation}\label{eq.inf.tr.var.cons.law}
\widetilde F^i=-Q_{(r)}F^i+(D_j\xi^i)F^j-(D_j\xi^j)F^i.
\end{equation}
Here $Q_{(r)}$ denotes the $r$-th prolongation of the operator $Q$.
\end{proposition}

Therefore, for any system of differential equations we can introduce a {\em generating set} of conservation laws in such a way
that all inequivalent conservation laws can be obtained from this set by
multiple using the symmetry transformations and taking linear combinations.
Such generating sets, called also $L$-bases, were firstly described in~\cite{Khamitova1982} (see also~\cite{Ibragimov1985}).

\begin{lemma}\label{PropositionOnInducedMapping}
Any point transformation $g$ between systems~$\mathcal{L}$ and~$\tilde{\mathcal{L}}$
induces a~linear one-to-one mapping $g_*$ between the corresponding linear spaces of conservation laws.
\end{lemma}

Consider the class~$\mathcal{L}|_S$ of systems~$L(x,u_{(\rho)},\theta(x,u_{(\rho)}))=0$
parameterized with the parameter-functions~$\theta=\theta(x,u_{(\rho)}).$
Here $L$ is a tuple of fixed functions of $x$, $u_{(\rho)}$ and $\theta$.
$\theta$ denotes the tuple of arbitrary (parametric) functions
$\theta(x,u_{(\rho)})=(\theta^1(x,u_{(\rho)}),\ldots,\theta^k(x,u_{(\rho)}))$
satisfying the additional condition~$S(x,u_{(\rho)},\theta_{(q)}(x,u_{(\rho)}))=0$.

Let~$P=P(L,S)$ denote the set of pairs each from which consists of
a system from~$\mathcal{L}|_S$ and a conservation law of this system.
Action of transformations from an equivalence group~$G^{\sim}$ of the class~$\mathcal{L}|_S$
together with the pure equivalence relation of conserved vectors
naturally generates an equivalence relation on~$P$.
Classification of conservation laws with respect to~$G^{\sim}$ will be understood as
classification in~$P$ with respect to the above equivalence relation.
This problem can be investigated in the way that it is similar to group classification in classes
of systems of differential equations. Specifically, we firstly construct the conservation laws
that are defined for all values of the arbitrary elements.
Then we classify, with respect to the equivalence group, arbitrary elements for each of the systems
that admits additional conservation laws.

For more detail and rigorous proof of the correctness of the above notions and statements see~\cite{Popovych&Ivanova2005CLs}.

\section{Classification of conservation laws}

To classify the conservation laws of equations from class~\eqref{eq_DifConvMultiDim}
we have to start our investigation from finding its equivalence transformations.
Application of both the direct and infinitesimal methods to class~\eqref{eq_DifConvMultiDim} allows us to
construct the complete equivalence group $G^{\Equiv}$.
The following statement is true.

\begin{theorem}
Any transformation from $G^{\mathop{\rm \, equiv}}$ has the
form
\begin{gather}\nonumber
\tilde t=\varepsilon_4t+\varepsilon_1, \quad
\tilde x^i=\varepsilon_5\mu_{ij}x^j+\varepsilon_{7i} t+\varepsilon_{2i}, \quad
\tilde u=\varepsilon_6u+\varepsilon_3, \\ \label{EquivTransf}
\tilde A=\varepsilon_4^{-1}\varepsilon_5^2A+\varepsilon_{8}, \quad
\tilde B^i=\varepsilon_4^{-1}\varepsilon_5\mu_{ij}B^j-\varepsilon_{7i}u+\varepsilon_{9i},
\end{gather}
where $\varepsilon$'s and $\mu$'s are constants,
$\varepsilon_4\varepsilon_5\varepsilon_6\ne0$, $M=(\mu_{ij})\in SO(n)$,
i.e. $M$ is an arbitrary special orthogonal matrix: $\det M=1$, $M^{\rm T}=M^{-1}$.
\end{theorem}

Any conservation law of equation from class~\eqref{eq_DifConvMultiDim} has the form
\begin{equation}\label{CLsEvolEq}
D_tT(t,x,u_{(r)})+D_iX^i(t,x,u_{(r)})=0,
\end{equation}
where~$T$ and~$X^i$ are often called {\em conserved density} and {\em conserved fluxes} correspondingly.
Since~\eqref{eq_DifConvMultiDim} is an evolution equation of the second order, then
up to the usual equivalence relation of conservation laws,
any conservation law of~\eqref{eq_DifConvMultiDim} is equivalent to one with the conserved density
$T=T(t,x,u)$ and the fluxes $X^i=X^i(t,x,u,u_1,\ldots,u_n)$~\cite{Ibragimov1985}.

We search (local) conservation laws of equations from class~\eqref{eq_DifConvMultiDim},
applying the modification of the direct method, which was proposed in~\cite{Popovych&Ivanova2005CLs}
and is based on using the notion of equivalence of conservation laws with respect to a transformation group.
First, expand total differentiation operators in~\eqref{CLsEvolEq} on the solution manifold of~\eqref{eq_DifConvMultiDim}:
\[
T_t+T_u(A_u u_{ii}+A_{uu}u_iu_i+B^i_uu_i)+X^i_i+X^i_uu_i+X^i_{u_j}u_{ij}=0,
\]
and split the obtained expression with respect to the unconstrained variables $u_{ij}$. Coefficients of $u_{ii}$ and $u_{ij}$, $i\ne j$ give
\[
X^i=-T_uA_uu_i+L^{[i,j]}_u(t,x,u)u_j+R^i(t,x,u),
\]
where $L^{[i,j]}(t,x,u)=-L^{[j,i]}(t,x,u)$.
Considering the conserved vector with the fluxes~$\tilde X^i=X^i-D_jL^{[i,j]}=-T_uA_uu_i+\tilde R^i(t,x,u)$,
which is equivalent to the initial one, without loss of generality we can assume that $L^{[i,j]}=0$.
Splitting the rest of the conservation law with respect to different powers of $u_i$ yields
\begin{gather*}
T_{uu}=0,\quad
-T_{iu}A_u+T_uB^i_u+R^i_u=0,\quad
T_t+R^i_i=0.
\end{gather*}
It follows that up to the usual equivalence relation of conserved vectors, the conserved density and fluxes in~\eqref{CLsEvolEq} have the form
\begin{equation}\label{CLsDensFlux}
T=\alpha(t,x)u,\qquad X^i=-\alpha A_uu_i+\alpha_iA-\alpha B^i,
\end{equation}
where $A,\, B^i$ and $\alpha(t,x)$ satisfy the classifying equation
\begin{equation}\label{eq_ClasCLs}
\alpha_tu+\Delta\alpha A-\alpha_iB^i=0.
\end{equation}

Let us note that function $\alpha(t,x)$ is a characteristics of conservation law~\eqref{CLsEvolEq},
and differentiating~\eqref{eq_ClasCLs} with respect to $u$ gives exactly
an equation for the adjoint symmetry for equation~\eqref{eq_DifConvMultiDim}.

Solving~\eqref{eq_ClasCLs} up to equivalence transformations~\eqref{EquivTransf} we obtain the complete classification of local conservation
laws of diffusion--convection equation~\eqref{eq_DifConvMultiDim}.

\begin{theorem}\label{TheoremCLsofDCEs}
A complete list of $G^{\Equiv}$-inequivalent equations~\eqref{eq_DifConvMultiDim} having
nontrivial conservation laws of form~\eqref{CLsEvolEq} with density and fluxes
\[
T=\alpha(t,x)u,\qquad X^i=-\alpha A_uu_i+\alpha_iA-\alpha B^i,
\]
is exhausted by the following:
\begin{gather*}
1.\quad \forall A, B^i: \qquad \alpha=1.\\
2.\quad \forall A,\ B^{1}=\ldots=B^k=0,\ B^{k+1}=A,\\
\phantom{2.\quad }{} B^{k+2},\ldots,B^n\not\in\langle 1,\, u,\, A \rangle\ \mbox{are linearly independent},\ 0\le k< n:\\
\phantom{2.\quad }{} \alpha=\alpha(x^1,\ldots,x^{k+1}),\
\alpha_{11}+\ldots+\alpha_{kk}+\alpha_{k+1,k+1}+\alpha_{k+1}=0.\\
3.\quad \forall A,\ B^1=\ldots=B^k=0,\ B^{k+1},\ldots,B^n\not\in\langle 1,\, u,\, A \rangle\ \mbox{are linearly independent},\\
\phantom{2.\quad }{}0\le k\le n:\qquad \alpha=\alpha(x^1,\ldots,x^{k}),\ \Delta\alpha=0.\\
4.\quad A=u, B^1=\ldots=B^k=0, \ B^{k+1}, \ldots, B^n\not\in\langle 1,\, u\rangle\  \mbox{are linearly independent},\\
\phantom{2.\quad}{}0\le k\le n:\qquad \alpha=\alpha(t,x_1,\ldots,x_k),\ \alpha_t+\Delta\alpha=0.
\end{gather*}
Together with values $A$ and $B^i$ we also adduce the constrains for characteristic $\alpha(t,x)$.
\end{theorem}
\begin{proof}
Let us briefly sketch the proof of the theorem. 
There exist two essentially inequivalent cases of integrating equation~\eqref{eq_ClasCLs}:
$A\in\langle 1,u\rangle$ and $A\not\in\langle 1,u\rangle$.

\noindent{\bf 1.} $A\not\in\langle 1,u\rangle$. Then, case $B^i\not\in\langle 1,u, A\rangle$, $1\le i\le n$ leads to the case~1 from the theorem list.
Suppose now that there exist such $j$ that $B^i=c^2_i A+c^1_i u+c^0_i$ for all $1\le i\le j$.
Then applying Galilean transformations $\tilde x^i=x^i+c^1_it$, $\tilde B^i=B^i-c^1_iu$ and translations of $B^i$
from the equivalence group, without loss of generality we set $c^1_i=c^0_i=0$. Thus, $B^i=c^2_iA\!\!\mod G^{\sim}$.
If all $c^2_i=0$ then we have
$B^i=0$, $1\le i\le j$ and classifying equation splits~\eqref{eq_ClasCLs} into the system $\alpha_t=\alpha_{i}=0$, $i\ge j+1$,
$\alpha_{ii}=0$. Thus, we have case~3. Suppose now, that there exists such $j_0$ that $c^{j_0}_2\ne0$.
Then up to the rotation transformations from the $G^{\sim}$ we can set $c^i_2=0$, for all $1\le i\le j-1$, $c^j=1$. Taking $j=k+1$ and decomposing
the classifying condition with respect to the linearly independent functions of $u$ we obtain precisely the case~2.

\noindent{\bf 2.} $A\in\langle 1,u\rangle$. Then $A=u\!\! \mod G^{\sim}$. If $B^i\not\in\langle 1,u\rangle$ for all $i$,
then $\alpha_i=0$, $\alpha_t=0$ that yields a subcase of case 1. If there exist $B^i\in\langle 1,u\rangle$ then,
applying the Galilean transformations and translations of $B^i$ we deduce that $B^i=0\!\! \mod G^{\sim}$ that leads to case~4.
\end{proof}

In the case of one space dimension $n=1$ equations for characteristic $\alpha$ can be integrated explicitly and we obtain simpler
form of classification result~\cite{Popovych&Ivanova2005CLs}.
\begin{corollary}
Any equation from class~\eqref{eq_DifConvMultiDim} with $n=1$ has a conservation law of form~\eqref{CLsEvolEq} where
\begin{equation}\textstyle\label{ker.cons.law}
1.\quad T=u,\qquad X=-A_uu_x-B.
\end{equation}
A complete list of $G^{\Equiv}$-inequivalent one-dimensional equations~\eqref{CLsEvolEq} having
additional (i.e. linear independent with~\eqref{ker.cons.law}) conservation laws
is exhausted by the following:
\begin{gather}\textstyle
2.\quad \forall A, \quad B=0: \qquad T=xu, \quad X=A-xA_uu_x,  \label{conslawB0}\\\textstyle
3.\quad \forall A, \quad B=A:
 \qquad T=e^xu, \quad X=-e^xA_uu_x,  \label{conslawBA}\\\textstyle
4.\quad A=u, \quad B=0: \qquad T=\alpha u, \quad X=\alpha_xu-\alpha u_x,  \label{conslawAuB0}
\end{gather}
where $x=x^1$, $X=X^1$, $B=B^1$,
$\alpha=\alpha(t,x)$ is an arbitrary solution of the backward linear heat equation $\alpha_t+\alpha_{xx}=0$.
(Together with values $A$ and $B$ we also adduce complete lists of
densities and fluxes of additional conservation laws.)
\end{corollary}

Analysis of the classification given in theorem~\ref{TheoremCLsofDCEs} and its corollary results in the following conclusions.
The space of conservation laws for equations from class~\eqref{eq_DifConvMultiDim} in general case is one-dimensional
and is generated by one with the conserved density $T=u$ and fluxes $X^i=-A_uu_i-B^i$.
Imposing restrictions on values of arbitrary functions $A(u)$ and $B^i(u)$ leads to subclasses of~\eqref{eq_DifConvMultiDim} with
two-dimensional spaces of (local) conservation laws for the nonlinear equation in one space dimension and with
infinite-dimensional spaces of conservation laws for linear equations and in the case $n\ge2$.
In this cases the space of conservation laws is parameterized with an arbitrary solution of a corresponding
adjoint linear equation.

\section{Composite variational principle for finding conservation laws}

Another possible way of finding conservation laws of equations~\eqref{eq_DifConvMultiDim} is extension of the Lagrangian approach via
introducing an `associate' equation in order to obtain a system of Euler--Lagrange equation
for some variational functional.
An example for such an associate is the adjoint symmetry equation~\cite{Atherton&Homsy1975,Vanberg1956}.
Here we illustrate this approach on example of subclass
\begin{equation} \label{eq_DifConvMultiDimB=0}
u_t=au_{ii}+a_uu_i^2,
\end{equation}
of equations~\eqref{eq_DifConvMultiDim} with zeroth convection terms $B^i=0$. Here and below $a=a(u)=A_u(u)$.

To find conservation laws of~\eqref{eq_DifConvMultiDimB=0} from the composite variational principle we consider system
\begin{gather}\label{sys_DifEqAdEc}
u_t=a_uu_i^2+au_{ii},\qquad v_t+av_{ii}=0
\end{gather}
obtained by combining equation~\eqref{eq_DifConvMultiDimB=0} and the adjoint equation
which is derivative of~\eqref{eq_ClasCLs} with respect to $u$.
This is a system of Euler--Lagrange equations for the variational functional
\begin{gather}\label{VarFunctionalDE}
\mathscr{L}=\int\left(\frac12(u_tv-uv_t)+au_iv_i\right)dx.
\end{gather}
with Lagrangian $\Lop=\frac12(u_tv-uv_t)+au_iv_i$. Indeed,
\begin{gather*}
{\Eop}^u\Lop=-\frac12v_t+a_uv_iu_i-\frac12v_t-D_i(av_i)=-(v_t+av_{ii}),\\[0.5ex]
{\Eop}^v\Lop=\frac12u_t+\frac12u_t+D_i(au_i)=u_t-(au_i+au_{ii}).
\end{gather*}
Thus, we can use the Noether theorem for finding conservation laws of system~\eqref{sys_DifEqAdEc}.

In this work we restrict ourselves to the investigation of conservation laws corresponding to the Lie point symmetries
of the Euler--Lagrange equations.
(The complete classification of conservation laws of~\eqref{sys_DifEqAdEc} requires finding of all its higher order symmetries.)

The Lie symmetries of system~\eqref{sys_DifEqAdEc} will be classified up to
the group~$\hat G^{\sim}$ of equivalence transformations
\begin{gather}\nonumber
\tilde t=\varepsilon_4t+\varepsilon_1, \quad
\tilde x^i=\varepsilon_5\mu_{ij}x^j+\varepsilon_{2i}, \quad
\tilde u=\varepsilon_6u+\varepsilon_3, \\ \label{EquivTransfSELE}
\tilde v=\epsilon v+\alpha(x),\quad
\tilde a=\varepsilon_4^{-1}\varepsilon_5^2a,
\end{gather}
where $\epsilon$, $\varepsilon$'s and $\mu$'s are constants,
$\varepsilon_4\varepsilon_5\varepsilon_6\epsilon\ne0$, $M=(\mu_{ij})\in SO(n)$,
i.e. $M$ is an arbitrary special orthogonal matrix: $\det M=1$, $M^{\rm T}=M^{-1}$,
$\alpha(x)$ is arbitrary solution of the Laplace equation $\alpha_{ii}=0$.
This group is a natural extension of transformations~\eqref{EquivTransf} (restricted to the case $B^i=0$)
onto the variable~$v$.

The classical Lie infinitesimal criterion yields the following system
\begin{gather*}
\tau_i=\tau_u=\tau_v=0, \quad \xi^i_u=\xi^i_v=0, \quad \eta_v=0, \quad \zeta_u=0,\\
\xi^i_j+\xi^j_i=0, \quad \eta_{uu}=\zeta_{vv}=0, \quad 2\xi^i_i-\tau_t=\frac{a_u}a\eta,\\
\eta_t-a\eta_{jj}=0, \quad \zeta_t+a\zeta_{jj}=0,\\
a(\xi^i_{jj}-2\eta_{iu})-2a_u\eta_i-\xi^i_t=0, \quad a(\xi^i_{jj}-2\zeta_{iv})+\xi^i_t=0.
\end{gather*}
of the determining equations for the coefficients of the infinitesimal symmetry generator
\[
Q=\tau(t,x,u,v)\p_t+\xi^i(t,x,u,v)\p_i+\eta(t,x,u,v)\p_u+\zeta(t,x,u,v)\p_v.
\]
Solving this system up to equivalence transformations~\eqref{EquivTransfSELE}
we obtain the complete group classification of system~\eqref{sys_DifEqAdEc}.

\begin{theorem}
The Lie algebra of the kernel of Lie invariance group of system~\eqref{sys_DifEqAdEc} is the algebra
\begin{gather*}
A^{\ker}=\langle \p_t,\; \p_i,\; x^i\p_j-x^j\p_i,\; \alpha(x)\p_v,\; v\p_v,\; 2t\p_t+x^i\p_i\rangle,
\end{gather*}
where $\alpha(x)$ is arbitrary solution of the Laplace equation $\alpha_{ii}=0$.

A complete set of $G^{\sim}$-inequivalent 
systems~\eqref{sys_DifEqAdEc} with the maximal Lie invariance algebras $A^{\max}\not=A^{\ker}$
is exhausted by the ones with following parameter-functions $a(u)$:
\begin{enumerate}
    \item $a=e^u$:\quad $A^{\max}=A^{\ker}+\langle t\p_t-\p_u\rangle$,
    \item $a=u^\mu$, ($\mu\ne0$):\qquad $A^{\max}=A^{\ker}+\langle \mu t\p_t-u\p_u \rangle$,
    \item $a=u^{-4/(n+2)}$, $n\ne2$: 
          \[A^{\max}=A^{\ker}+\langle 4t\p_t+(n+2)u\p_u,\;
          (2x^ix^j-\delta^{ij}|x|^2)\p_j-(n+2)x^iu\p_u-(1+\mu)(n+2)x^iv\p_v \rangle,\]
    $a=u^{-1}$, $n=2$: \quad $A^{\max}=A^{\ker}+\langle t\p_t+u\p_u,\; \phi(x^1,x^2)\p_1+\psi(x^1,x^2)\p_2-2\phi_1(x^1,x^2)u\p_u\rangle$,\\
    where $\phi(x^1,x^2)$ and $\psi(x^1,x^2)$ are arbitrary solutions of the Cauchy--Riemann  system $\phi_1=\psi_2$, $\phi_2=-\psi_1$,
    \item $a=1$:
   \begin{gather*}
    A^{\max}=\langle \p_t,\; \p_i,\; x^i\p_j-x^j\p_i,\; \alpha(t,x)\p_v,\; v\p_v,\; 2t\p_t+x^i\p_i,\; u\p_u,\; 2t\p_i-x^iu\p_u+x^iv\p_v,\; \\
   \phantom{A^{\max}=}{}    4t^2\p_t+4tx^i\p_i-(|x|^2+2nt)u\p_u+(|x|^2-2nt)v\p_v,\; \beta(t,x)\p_u\rangle,
   \end{gather*}
   where $\alpha(t,x)$, $\beta(t,x)$ are arbitrary solutions of the linear heat equations
$\alpha_t+\alpha_{ii}=0$ and $\beta_t=\beta_{ii}$.
\end{enumerate}
\end{theorem}

It is the obvious extension of the known symmetry algebras~\cite{Dorodnitsyn&Knyazeva&Svirshchevskii1983,Dorodnitsyn&Svirshchevskii1983,Ibragimov1994V1}
of diffusion equations to the adjoint equation.


Applying the infinitesimal criterion for the variational invariance we get the basis
of the algebra of variational symmetries
\[
A^{\ker}_{\rm var}=\langle P_t=\p_t,\; P_i=\p_i,\; J_{ij}=x^i\p_j-x^j\p_i,\; V(\alpha(x))=\alpha(x)\p_v,\; D_1=2t\p_t+x^i\p_i-nv\p_v\rangle,
\]
of variational functional~\eqref{VarFunctionalDE} for arbitrary $a(u)$ and its extension
\begin{gather*}
A^{1}_{\rm var}=\langle P_t=\p_t,\; P_i=\p_i,\; J_{ij}=x^i\p_j-x^j\p_i,\; V_1(\alpha(t,x))=\alpha(t,x)\p_v,\\
\phantom{A^{\max}=}{} D_1=2t\p_t+x^i\p_i-nv\p_v,\; D_2=u\p_u-v\p_v,\; G_i=2t\p_i-x^iu\p_u+x^iv\p_v,\; \\
\phantom{A^{\max}=}{}\Pi=4t^2\p_t+4tx^i\p_i-(|x|^2+2nt)u\p_u+(|x|^2-2nt)v\p_v,\; U(\beta(t,x))=\beta(t,x)\p_u, \rangle
\end{gather*}
for $a(u)=1$.
For $a(u)=e^u$, $A_{\rm var}=A^{\ker}_{\rm var}+\langle D_2=t\p_t-\p_u\rangle$.

If $a(u)=u^{\mu}$, $\mu\ne0,-4/(n+2)$ then $A_{\rm var}=A^{\ker}_{\rm var}+\langle D_2=\mu t\p_t-u\p_u+v\p_v\rangle$.

For $a=u^{-4/(n+2)}$,
\begin{gather*}
A_{\rm var}=A^{\ker}+\langle D_2=4t\p_t+(n+2)u\p_u,\\
\phantom{A_{\rm var}=}          \Pi_2=(2x^ix^j-\delta^{ij}|x|^2)\p_j-(n+2)x^iu\p_u-(1+\mu)(n+2)x^iv\p_v \rangle.
\end{gather*}

According to the generalized Noether theorem there exists a one-to-one correspondence between the found variational symmetries
and the nontrivial conservation laws of system~\eqref{sys_DifEqAdEc}.
Thus, using the well-known formulae for the components of the conserved vectors (see, e.g.,~\cite[Corollary~4.30]{Olver1986})
we obtain nontrivial conservation laws of system~\eqref{sys_DifEqAdEc} presented in table 1.


{\setcounter{table}{0}\setcounter{tbn}{0}
\begin{center}\renewcommand{\arraystretch}{1.1}\refstepcounter{table}
Table~\thetable:  Conserved vectors of nonlinear system~\eqref{sys_DifEqAdEc},\\[2ex]\footnotesize
\begin{tabular}{|l|l|l|l|}
\hline \vspacebefore
N &\hfill$Q \hfill$ & \hfill $T \hfill$ &  \hfill $X^i \hfill$\\[0.5ex]
\hline \vspacebefore
\refstepcounter{tbn}\thetbn& $V(\alpha) $ & $\alpha(x)u$ & $\alpha_iA-\alpha au_i$\\[0.5ex]
\refstepcounter{tbn}\thetbn& $P_t$ & $au_iv_i$ & $-a(u_tv_i+u_iv_t)$\\[0.5ex]
\refstepcounter{tbn}\thetbn& $P_j$ & $\frac12(uv_j-vu_j)$ & $\delta^{ij}\left(\frac12(u_tv-uv_t)+au_jv_j\right)-au_iv_j-au_jv_i$\\[0.5ex]
\refstepcounter{tbn}\thetbn& $J_{kl}$ & $\frac12v(u_k-u_l)+\frac12u(v_l-v_k) $ & $(u_k-u_l)av_i+(v_k-v_l)au_i $\\[0.5ex]
\refstepcounter{tbn}\thetbn& $D_1$ & $2tau_jv_j+\frac12x^j(uv_j-u_jv)+\frac{n}{2}uv$ &
$\frac12x^i(u_tv-uv_t)-av_i(x^ju_j+2tu_t)-au_i(nv+x^jv_j+2tv_t)+$\\ &&& $+ax^iu_jv_j $\\[0.5ex]
\hline
\end{tabular}
\end{center}}
{\noindent\footnotesize
Here $A=A(u)=\int a(u)du$, $\alpha=\alpha(x)$ is an arbitrary solution of the Laplace equation $\alpha_{ii}=0$,
$\delta^{ij}=1$ if $i=j$ and $\delta^{ij}=0$ if $i\ne j$, $1\le i,j,k,l\le n$.
}
\bigskip

System~\eqref{sys_DifEqAdEc} with $a=e^u$ admits extra conservation law of form $D_t(te^uu_iv_i-v)-D_i(t(u_tv_i+u_iv_t)+av_i)=0$.

For case $a=u^\mu$, $\mu\ne0,-4/(n+2)$, system~\eqref{sys_DifEqAdEc} has a CL with
the conserved density and fluxes of form $T=\mu tau_iv_i-uv$, $X^i=-a(uv_i-u_iv)-\mu t a(u_tv_i+u_iv_t)$.
If $\mu=-4/(n+2)$ system~\eqref{sys_DifEqAdEc} admits additionally
\begin{gather*}
D_t\left[-\dfrac12(2x^ix^j-\delta^{ij}|x|^2)(u_jv-uv_j)-\mu(n+2)x^iv\right]\\
+D_x\bigg[(n+2)\delta^{ij}u^{\frac{n-2}{n+2}}v+(2x^ix^j-\delta^{ij}|x|^2)\left(\dfrac12(u_tv-uv_t)+a(u_kv_k-u_jv_i-u_iv_j)\right)\\
-(n+2)x^ia(uv_i-(1+\mu)u_iv)\bigg]=0.
\end{gather*}

Conservation laws of the linear system are presented in table~2.

{\setcounter{tbn}{0}
\begin{center}\renewcommand{\arraystretch}{1.1}\refstepcounter{table}
Table~\thetable:  Conserved vectors of system $u_t-u_{ii}=0,\ v_t+v_{ii}=0$\\[2ex]\footnotesize
\begin{tabular}{|l|l|l|l|}
\hline \vspacebefore
N &\hfill$Q \hfill$ & \hfill $T \hfill$ &  \hfill $X^i \hfill$\\[0.5ex]
\hline \vspacebefore
\refstepcounter{tbn}\thetbn\label{CLLinP_v^1}\hspace{-1mm}& $V_1(\alpha)$ & $\alpha u$ & $\alpha_iu-\alpha u_i$\\[0.5ex]
\refstepcounter{tbn}\thetbn\label{CLLinP_t}\hspace{-1mm}& $P_t$ & $u_iv_i$ & $-(u_tv_i+u_iv_t)$\\[0.5ex]
\refstepcounter{tbn}\thetbn\label{CLLinP_j}\hspace{-1mm}& $P_j$ & $\frac12(uv_j-vu_j)$
 & $\delta^{ij}\left(\frac12(u_tv-uv_t)+u_jv_j\right)-u_iv_j-u_jv_i$\\[0.5ex]
\refstepcounter{tbn}\thetbn\label{CLLinJ}\hspace{-1mm}& $J_{kl}$ & $\frac12v(u_k-u_l)+\frac12u(v_l-v_k) $ & $(u_k-u_l)v_i+(v_k-v_l)u_i $\\[0.5ex]
\refstepcounter{tbn}\thetbn\label{CLLinD_1}\hspace{-1mm}& $D_1$ & $2tau_jv_j+\frac12x^j(uv_j-u_jv)+\frac{n}{2}uv$ &
$\frac12x^i(u_tv-uv_t)-v_i(x^ju_j+2tu_t)-u_i(nv+x^jv_j+2tv_t)+$\\ &&& $+x^iu_jv_j $\\[0.5ex]
\refstepcounter{tbn}\thetbn\label{CLLinP_u^1}\hspace{-1mm}& $U(\beta)$ & $\beta v$ & $\beta v_i-\beta_iv $\\[0.5ex]
\refstepcounter{tbn}\thetbn\label{CLLinD_2}\hspace{-1mm}& $D_2$ & $uv$ & $uv_i-u_iv $\\[0.5ex]
\refstepcounter{tbn}\thetbn\label{CLLinG_j}\hspace{-1mm}& $G_j$ &$-x^juv-t(u_jv+uv_j)$
& $\delta^{ij}\left(\frac12(vu_t-uv_t)+au_iv_i\right)-x^j(uv_i+u_iv)-2t(u_jv_i+u_iv_j)$\\[0.5ex]
\refstepcounter{tbn}\thetbn\label{CLLinPi}\hspace{-1mm}& $\Pi$ &$4t^2u_iv_i+2tx^i(uv_i-u_iv)-|x|^2uv$ &
$2tx^i(u_tv-uv_t)+4tx^iu_jv_j-4tx^i(u_jv_i+u_iv_j)-$\\[0.5ex]
& && $-uv_i(|x|^2+2nt)+u_iv(|x|^2-2nt)-4t^2(u_tv_i+u_iv_t) $\\
\hline
\end{tabular}
\end{center}}
{\noindent\footnotesize
Here  $\alpha=\alpha(t,x)$, $\beta=\beta(t,x)$ are arbitrary solutions of the linear heat equations
$\alpha_t+\alpha_{ii}=0$ and $\beta_t=\beta_{ii}$, $\delta^{ij}=1$ if $i=j$ and $\delta^{ij}=1$ if $i\ne j$.
}
\bigskip

Note that the conserved vectors from case~1 of tables~1--2 do not involve the solutions of the adjoint equation,
and therefore, determine local conservation laws of diffusion equation~\eqref{eq_DifConvMultiDimB=0}.
These are exactly the conservation laws obtained above with application of the direct method (see theorem~\ref{TheoremCLsofDCEs}).
The conserved vectors from cases~1--5 (1--9) of table~1 (table~2) determine the nonlocal conservation laws
of diffusion equation~\eqref{eq_DifConvMultiDimB=0},
where the non-local variable $v$ is defined by the associated equation $v_t+av_{ii}=0$.

Consider now in more detail the linear case
\begin{gather}\label{sys_DifEqAdEcLinear}
u_t-u_{ii}=0,\qquad v_t+v_{ii}=0
\end{gather}
Let us emphasize that the conserved vectors presented in table~2 are inequivalent
with respect to the usual equivalence relation given in definition~\ref{DefinitionOfConsVectorEquivalence}
and the corresponding conservation laws are linearly independent.
At the same time, acting by the transformations from Lie symmetry group of system~\eqref{sys_DifEqAdEcLinear},
we can establish $G^{1}_{\rm var}$-equivalence relations on the space of conservation laws.
Thus, case~\ref{CLLinP_u^1} is equivalent to case~\ref{CLLinP_v^1} under the discrete symmetry transformation
$\tilde t=-t$, $\tilde x=x$, $\tilde u=-v$, $\tilde v=-u$.

Consider now the set of conservation laws spanned by those from cases~\ref{CLLinP_v^1} and~\ref{CLLinD_2}.
It is obvious that they form a linear subspace of the space of conservation laws of system~\eqref{sys_DifEqAdEcLinear}
with a set of generatrices $\{(\alpha u,\alpha_iu-\alpha u_i),\;(uv,uv_i-u_iv)\}$
(as usual, we identify conservation laws with their representatives in the space of conserved vectors).
Let us now change the set of generatrices as $\{((v+\alpha u),(v_i+\alpha_i)u-(v+\alpha)u_i),\;(uv,uv_i-u_iv)\}$.
Under the point transformation $v\to v-\alpha$ (the other variables are not transformed) from the symmetry group~$G^{1}_{\rm var}$
the conserved vector $((v+\alpha u),(v_i+\alpha_i)u-(v+\alpha)u_i)$ is mapped to the $(uv,uv_i-u_iv)$.
Therefore, up to the symmetry group of system~\eqref{sys_DifEqAdEcLinear} the considered subspace of conservation laws is generated
by the conservation law with the conserved vector $(uv,uv_i-u_iv)$.

In~\cite{Ibragimov1985,Khamitova1982} it is shown that for a system of Euler--Lagrange equations
a generating set of inequivalent conservation laws is determined by the ones with conserved vectors
that correspond to symmetry operators spanning the symmetry algebra under the action of adjoint representations.
Using this property we can complete investigation of equivalences between the conservation laws of system~\eqref{sys_DifEqAdEcLinear}.
Since $[P_1,G_1]=D_2$, $[P_t,D_1]=2P_t$ and $[P_t,\Pi]=4D_1-2nD_2$, the conservation laws corresponding to $P_t$, $D_1$ and $D_2$ are dependent
on those generated by $P_1$, $G_1$, and $\Pi$.
Using that $[P_t,G_j]=P_j$ we deduce that the conservation law generated by $P_j$ can be obtained from the one corresponding to $G_j$
under the multiple actions of symmetry transformations.
It is obvious, that any conservation law from case~\ref{CLLinJ} is equivalent to those generated by $J_{12}$ with respect to
the rotation transformations. Similarly, case~\ref{CLLinG_j} is $G^{1}_{\rm var}$-equivalent to the conservation law generated by $G_1$.
Therefore, the following statement is true.

\begin{theorem}
The generating set of conservation laws of system~\eqref{sys_DifEqAdEcLinear} consists of
the conservation laws corresponding to the variational symmetry operators $J_{12}$, $G_1$, $\Pi$.
\end{theorem}

Analogously we can find the generating set of conservation laws of nonlinear system~\eqref{sys_DifEqAdEc}.

\begin{theorem}
The generating set of conservation laws of nonlinear system~\eqref{sys_DifEqAdEcLinear} with arbitrary value of $a(u)$ consists of
the conservation laws corresponding to the variational symmetry operators $D_1$, $J_{12}$ and $V(\alpha(x))$.
For $a=e^u$ and $a=u^\mu$ all the conservation laws are generated by ones corresponding to $D_1$, $D_2$, $J_{12}$ and $V(\alpha(x))$.
\end{theorem}

\begin{note}
There exist a non-trivial $G^{\max}$-equivalence relation on the set of conservation laws of system~\eqref{sys_DifEqAdEcLinear}
corresponding to the variational symmetry operators $V(\alpha(x))$,
generated by the rotation transformations, translations of space variables and scale transformations.
\end{note}

\section{Conclusion}

In the presented work we used the direct method and the composite variational principle for
construction of all possible inequivalent local conservation laws for
class~\eqref{eq_DifConvMultiDim} of $n$-dimensional diffusion--convection equations.
In contrast to the case of one-dimensional diffusion--convection equations, there exist multidimensional nonlinear equations with
infinite-dimensional spaces of conservation laws.
The composite variational principle allows us to determine a set of nonlocal conservation laws for the equations under consideration.

The adduced results can be developed and generalized in a number of directions.
Thus, e.g., the problem of the complete classification of conservation laws of the obtained system of Euler--Lagrange equations,
which is closely related to construction of generalized (higher order) symmetry operators, is still open.
Another promising direction is connected with investigation of potential symmetries and potential conservation laws.
For such systems (see e.g.~\cite{Anco&Bluman1997}) it follows from existence of gauge transformations for potentials
that there exist no nontrivial potential symmetries without the introduction of a gauge.
A natural question arises: How does one choose a gauge to obtain a nontrivial potential symmetry?
(Existence of potential symmetries can be used as a criterion of adequacy of gauges
applied in physical theories.) Moreover, we should like to classify gauges depending on group
properties of potential symmetries up to symmetry group of the system under consideration.
When potentials are introduced and a gauge is chosen, we can investigate symmetries of
the potential system, construct its conservation laws and introduce new additional potential variables.
For the extended potential system we again have a problem of choosing a
gauge. Iterating the above procedure, we come to the problem of finding a hierarchy of right
gauges.

\subsection*{Acknowledgements}
The author is grateful to G.~Bluman and R.O.~Popovych for useful comments and fruitful discussion.
She also acknowledges financial support from the National Sciences and Engineering Research Council of Canada and
Department of Mathematics of the University of British Columbia.
This research was partially supported by the grant of the President of Ukraine for young scientists (project number GP/F11/0061Â).


\begin{thebibliography}{99}
\footnotesize
\itemsep=0.2ex

\bibitem{Anco&Bluman1997}
Anco S.C. and Bluman G.,
Nonlocal symmetries and nonlocal conservation laws of Maxwells equations,
{\it J.~Math. Phys.}, 1997, V.38, 3508--3532.

\bibitem{Anco&Bluman2002a}
Anco~S.C. and Bluman~G.,
Direct construction method for conservation laws of partial differential equations.~I.
Examples of conservation law classifications,
{\it Eur. J. Appl. Math.}, 2002, V.13, Part 5, 545--566 (math-ph/0108023).

\bibitem{Anco&Bluman2002b}
Anco~S.C. and Bluman~G.,
Direct construction method for conservation laws of partial differential equations.~II. General treatment,
{\it Eur. J. Appl. Math.}, 2002, V.13, Part 5, 567--585 (math-ph/0108024).


\bibitem{Atherton&Homsy1975}
Atherton R.W., Homsy G.M.
On the existence and formulation of variational principles for nonlinear differential equations,
{\it Stud. App. Math.}, 1975, V.54, 31--60.

\bibitem{Blaszak1998}
B\l aczak M.,
Multi-Hamiltonian theory of of dynamical systems, Berlin, Springer, 1998.


\bibitem{Chayes&Osher&Ralston1993}
Chayes J.T., Osher S.J. and Ralston J.V.,
On singular diffusion equations with applications to self-organized criticality,
{\it Comm. Pure Appl. Math.}, 1993, V.46, 1363--1377.

\bibitem{Cherniha&King2005}
Cherniha R. and King J.R.,
Lie symmetries and conservation laws of non-linear multidimensional reaction–diffusion systems with variable diffusivities,
{\it IMA J. Appl. Math.}, 2005, Advance Access published on September 23.

\bibitem{Dorodnitsyn&Knyazeva&Svirshchevskii1983}
Dorodnitsyn V.A., Knyazeva I.V. and Svirshchevski\u\i S.R.,
Group properties of the anisotropic heat equation with source $T\sb{t}=\sum \sb{i}(K\sb{i}(T)T\sb{x\sb{i}})\sb{x\sb{i}}+Q(T)$,
Preprint N~134, Akad. Nauk SSSR, Inst. Prikl. Mat., 1982.

\bibitem{Dorodnitsyn&Svirshchevskii1983}
Dorodnitsyn~V.A and Svirshchevskii~S.R.,
On Lie--B\"acklund groups admitted by the heat equation with a~source,
Preprint N~101, Moscow,  Keldysh Institute of Applied Mathematics of Academy of Sciences USSR, 1983.

\bibitem{Fushchych&Nikitin1994}
Fushchich~W.I., Nikitin~A.G.,
Symmetries of Equations of Quantum Mechanics, New York, Allerton Press Inc., 1994.

\bibitem{Gennes1983}
De Gennes P.G.,
Wetting: statics and dynamics,
{\it Reviews of Modern Physics}, 1985, V. 57, 827--863.

\bibitem{Ibragimov1985}
Ibragimov~N.H.,
Transformation groups applied to mathematical physics,
{\it Mathematics and its Applications (Soviet Series)}, Dordrecht, D. Reidel Publishing Co., 1985.

\bibitem{Ibragimov1994V1}
Ibragimov~N.H. (Editor),
Lie group analysis of differential equations -- symmetries, exact solutions and conservation laws,
Vol.~1, Boca Raton, FL, Chemical Rubber Company, 1994.

\bibitem{Ibragimov&Kolsrud2004}
Ibragimov~N.H. and Kolsrud T.,
Lagrangian approach to evolution equations: symmetries and conservation laws,
{\it Nonlinear Dynamics}, 2004, V.36, 29--40.

\bibitem{Ivanova2004}
Ivanova N.,
Conservation laws and potential systems of diffusion-convection equations,
{\it Proceedings of Fifth International Conference ``Symmetry in Nonlinear Mathematical Physics''}
(23--29 June, 2003, Kyiv),
Kyiv, Institute of Mathematics, 2004, Part~1, 149--153 (math-ph/0404025).

\bibitem{Ivanova&Popovych&Sophocleous2004}
Ivanova~N.M., Popovych~R.O. and Sophocleous~C.,
Conservation laws of variable coefficient diffusion--convection equations,
{\it Proceedings of Tenth International Conference in Modern Group Analysis}, (Larnaca, Cyprus, 2004), 107--113 (math-ph/0505015).

\bibitem{Kara&Mahomed2002}
Kara~A.H. and Mahomed~F.M.,
A basis of conservation laws for partial differential equations,
{\it J. Nonlinear Math. Phys.}, 2002, V.9, 60--72.

\bibitem{Khamitova1982}
Khamitova R.S.,
The structure of a group and the basis of conservation laws,
{\it Teoret. Mat. Fiz.}, 1982, V.52, N~2, 244--251.

\bibitem{Olver1986}
Olver P., Applications of Lie groups to differential equations,
New-York, Springer-Verlag, 1986.

\bibitem{Popovych&Ivanova2005CLs}
Popovych R.O. and Ivanova N.M.,
Hierarchy of conservation laws of diffusion--convection equations,
{\it J. Math. Phys.}, 2005, V.46, 043502 (math-ph/0407008).

\bibitem{Richard's1931}
Richard's L.A.,
Capillary conduction of liquids through porous mediums,
{\it Physics}, 1931, V.1, 318–333.

\bibitem{Romanovsky1989}
Romanovsky Y.R.,
On symmetries of the heat equation,
{\it Acta Appl. Math.}, 1989, V.15, 149--160.

\bibitem{Vanberg1956}
Vainberg M.M.,
Variational methods for investigation of non-linear operators (Russian),
{\it Gosudarstv. Izdat. Tehn-Teor. Lit.}, Moscow, 1956;
English translation: {\it Holden-Day}, San Francisco, Calif., 1964.

\bibitem{Wolf2002}
Wolf~T.
A comparison of four approaches to the calculation of conservation laws,
{\it Eur. J. Appl. Math.}, 2002, V.13, Part 5, 129--152.

\end{thebibliography}
\end{document}